\documentclass[pre,aps,twocolumn,showpacs]{revtex4}
\usepackage{graphicx}

\begin{document}

\title{Graphene nanopore devices for DNA sequencing: A tight-binding model study}

\author{Sourav Kundu}

\email{phys.abcd@protonmail.com}

\affiliation{Condensed Matter Physics Division, 
Saha Institute of Nuclear Physics, 
1/AF, Bidhannagar, Kolkata 700 064, India}

\author{S. N. Karmakar}

\affiliation{Condensed Matter Physics Division, 
Saha Institute of Nuclear Physics, 
1/AF, Bidhannagar, Kolkata 700 064, India}

\begin{abstract} 
  We present a tight-binding model study of a two-terminal graphene nanopore device for sequential determination of DNA bases. Using Green's function technique we investigate the changes in electronic transport properties of the device due to insertion of different nucleotides into the nanopore created within a zigzag graphene nanoribbon. First we try to characterise the device in static condition and then go for sequencing application by setting the bias across it to a specific voltage and then recording the characteristic current signals corresponding to each nucleotides of a translocating DNA. Our investigations show that graphene nanopores can certainly become very efficient and reliable for sequencing applications in future.
  
\end{abstract} 

%\begin{keyword}
 % Graphene nanopore \sep DNA sequencing \sep Biosensor
  
%  \PACS 73.23.-b \sep 73.63.-b \sep 87.14.gk \sep 87.15.Pc
 %\end{keyword}

\maketitle

\section{Introduction:}

The idea of nanopore based DNA sequencing has emerged from the need to diminish the cost and increase the speed of genome sequencing. Traditional marker based DNA sequencing needs much more time and also quite costly as all the biomolecules targeted for sequential determination has to undergo fluorescent labeling~\cite{jett,sauer,werner}. Dramatic advancements in the DNA sequencing techniques in the past decade have huge impact on genome research~\cite{eid,clarke,manzo,wells,osaki}. As the understanding of genetic components of different diseases now become more vivid, it leads to the development of new type of medical treatments and diagnostics. Genetic research has many other applications apart from human medicine, such as in security, biology and agriculture. But the main problem with the current sequencing technology {\it e.g.}, the well-established Sanger method~\cite{sanger}, is its cost. Sequencing a specific human genome to high quality currently costs around \textdollar1K depending on the procedures. Whereas the nanopore based DNA sequencing offers marker-free single-molecule detection without any sample modification at much lower cost. The initial attempts of nanopore based sequencing of single-stranded DNA  (ss-DNA) using protein pore ($\alpha$-Haemolysin) were made more than two decades ago by translocating ss-DNA through the nanopore under an applied bias while reading the changes in the ionic current passing through the pore~\cite{bayley,branton,kasiano,deamer}. Detection of the signatures of the nucleotides trapped inside a nanopore {\it i.e.}, of a static strand in nanopore is easy~\cite{ashken}, whereas the problem of detecting the responses of the nucleotides of a translocating DNA through nanopore is still a big challenge, as the speed of translocation is too high for getting the necessary current resolution unless they are slowed down by chemical modifications with bulky groups~\cite{mitchell}. The early usage of biological nanopore in DNA sequencing is mainly due to the key factors like nanometer dimension of the pore, detectable ion-current modification, and abundance of different processing enzymes and chemicals to regulate the speed of DNA translocation in discrete steps~\cite{cherf,manrao}. But there are some disadvantages with biological nanopores due to their poor mechanical strength~\cite{striemer}, and difficulty in integrating them with on-chip electronics~\cite{Rosen}. However, Oxford Nanopore Technologies has been able to commercialize their products based on biological nanopores with much success. Solid-state nanopores overcome these drawbacks, and also provide some additional advantages like possibility of other types of detection, different from ionic current~\cite{lagerq,zwolak,gracheva,sigalov,mcnally,huang,tsutsui,xie,fyta}. But one big problem remains, the synthetic membranes used for detection are more than 10 nm thick, which means that several nucleotides will occupy the nanopore at the same time (distance between two consecutive nucleotides is 0.34 nm in double-stranded DNA and 0.5 nm in ss-DNA~\cite{ralph,dekker}) and it becomes impossible to perform single-molecule base-specific measurements. Graphene, a single layer allotrope of carbon~\cite{novo1}, provides a new way to develop nanopore based technique for the single-molecule sequencing of DNA. Graphene is much more superior over the conventional solid-state nanopores as the nanopore dimension can be controlled to the atomic length scale as well as the thickness of the membrane can be of the order of the distance between two consecutive nucleotides in a DNA chain. Graphene shows better electromechanical properties~\cite{rocha}. Graphene based nanopore also offers several ways of sequential detection by means of transverse tunnelling~\cite{postma,prasong}, nanoribbon conductance~\cite{saha,nelson,sourav4} and also using mechanical deformation of DNA~\cite{sathe} followed by some experimental realizations~\cite{traversi,menard}.

 In the present work we report a tight-binding formulation based investigation of a graphene nanopore 
sequencing device. Though there are several reports~\cite{prasong,saha,he,venkat,chien,pathak,prasong1} available in the literature that address nanopore sequencing of DNA, there reamins ample scope to explore thses systems in details. Here we present a detail physical understanding about the nanopore sequencing devices that are based on transverse electronic transport using tight-binding approach. Using Green's function approach and Landauer formalism we study the local density of states (LDOS), conductance and I-V response to detect the sequential arrangement of the nucleotides while trapping them inside the nanopore at static condition as well as by translocating a ss-DNA through the nanopore. We report the performance of the device on the basis of our model calculations and show that the characteristic current amplitudes for the four different nucleotides are quite distinct so that they could be detected unambiguously. The present study reveals the graphene based nanopore devices could be used for reliable sequential detection of the DNA bases in near future.

\section{Theoretical Formulation} 

\begin{figure}[ht]
  \centering

    \includegraphics[width=68mm,height=47mm]{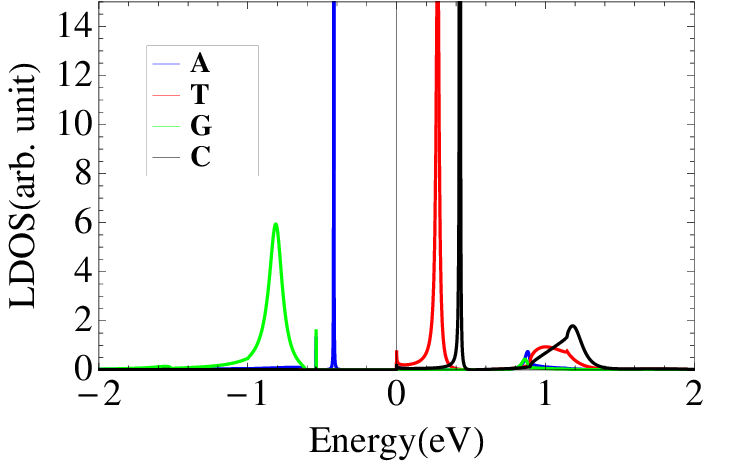}

\caption{(Color online). LDOS of the four nucleotides trapped at 
the nanopore. There are four distinct peaks representing different 
bases close to their respective characteristic site-energies.}
\label{fig1}
\end{figure}

 In order to determine the sequence of the nucleotides we use 
a graphene nanoribbon with zigzag edges with a pore at its 
centre (Fig.~\ref{fig1}). We create the pore in such a way that 
the bipartite two-sublattice nature of graphene is maintained. 
We use semi-infinite zigzag graphene nanoribbon (ZGNR) as 
electrodes~\cite{saha} also, attached to the both sides of the device. 
The active ZGNR system including the nanopore (excluding the electrodes) 
termed as device can be represented by the following tight-binding 
Hamiltonian 
\begin{eqnarray}
 &H_{zgnr}&= \sum\limits_{i=1}^N\left(\epsilon
c^\dagger_{i}c_{i}+tc^\dagger_{i}c_{i+1}+\mbox{H.c.} \right)
\end{eqnarray}
%%%%%%%%%%%%%%%%%%%%%%%%%%%%%% fig1

 where $\epsilon$ is the site-energy of each carbon atom in ZGNR and $t$ is the nearest neighbour hopping amplitude. $c_i$ 
and $c^\dagger_{i}$ represent electron annihilation and creation operators at the ith Wannier state respectively. The Hamiltonian of the ZGNR electrodes can also be expressed in the same way. The total Hamiltonian of the entire system can be written as $H_{tot}=H_{zgnr}+ H_{electrodes}+H_{tun1}~,\label{hamilton}$ where $H_{tun1} = \tau \left(c^\dagger_0c_1+c^\dagger_Nc_{N+1} +\mbox{H.c.}\right)$ is the coupling Hamiltonian between the active ZGNR device and the electrodes and the coupling between them is defined by the hopping integral $\tau$. %$H_{tun2} = \tau' \left(c'^\dagger_0c'_1+c'^\dagger_Nc'_{N+1} +%\mbox{H.c.}\right)$ is the coupling Hamiltonian between edge atoms of the zgnr nanopore and DNA bases. 
In our calculations we set $t$=1.0 eV and $\epsilon$=0 eV, and $\tau$=$t$ as both the nanopore device and electrodes are zgnr. The Hamiltonian of a ss-DNA chain is given by 
\begin{eqnarray}
& H_{DNA}&= \sum\limits_{i=1}^N\left(\epsilon_{i}
c^\dagger_{i}c_{i}+t_{i,i+1}c^\dagger_{i}c_{i+1}+\mbox{H.c.} \right)
\end{eqnarray}
where $\epsilon_{i}$ is the site energy of the respective nucleotides and $t_{i,i+1}$ is the hopping amplitude between them. The $H_{DNA}$ is coupled with the active device ($H_{zgnr}$) via a coupling parameter $\tau'$ where a nucleotide of a translocating DNA interacts with the edge atoms of the 
nanopore created within the zgnr.

The LDOS profiles of the nucleotides trapped at the nanopore are defined in terms of the Green's function as $\rho(E,i) = - \frac{1}{\pi} {\rm Im[G_{ii}(E)]}$
where, $G(E)= (E-H+i\eta)^{-1}$ is the Green's function for the entire system including the nucleotides with electron energy $E$ as $\eta\rightarrow0^+$, $H=$ Hamiltonian of the entire system, and, ${\rm Im}$ represents the imaginary part. $\rm G_{ii}(E)$ is the diagonal matrix element ($<i|G(E)|i>$) of the Green's function, $|i>$ being the Wannier state associated with the trapped nucleotide. To obtain the transmission probability of an electron, we first calculate the self-energy of the left (right) semi-infinite zgnr electrode $\Sigma_{L(R)}$ following the recursive Green's function technique~\cite{nardelli,lopez}. 
The single-particle retarded Green's function for the entire system at an energy $E$ is given by $G^r=[G^a]^\dagger=[E- H_{zgnr}-\Sigma^r_L-\Sigma^r_R+i\eta]^{-1}$, where $\Sigma^{r(a)}_{L(R)}=H^\dagger_{\mbox{tun}} G^{r(a)}_{L(R)} H_{\mbox{tun}}$ is the retarded (advanced) self-energy of the left (right) electrode and $G^{r(a)}_{L(R)}$ is the retarded (advanced) Green's function for the left (right) electrode. The transmission probability of an electron with incident energy $E$ is given by 
$T(E)={\mbox {\rm Tr}} [\Gamma_L G^r \Gamma_R G^a]$, where the trace is over the entire Hilbert space spanned by the zgnr and the DNA nucleotide and $\Gamma_{L(R)}=i[\Sigma^r_{L(R)}-\Sigma^a_{L(R)}]$ is the linewidth function. At absolute zero temperature, in the Landauer formalism current 
through the system for an applied bias voltage V is given by
$I(V)=\frac{2e}{h} \int^{E_F+eV/2}_{E_F-eV/2} T(E)dE~$, where $E_F$ is the Fermi energy of the system. 
We have assumed that there is no charge accumulation within the system.

\section{Results:}

%%%%%%%%%%%%%%%%%%%%%%%%%% fig2
\begin{figure}[ht]
 \centering

    \includegraphics[width=68mm,height=52mm]{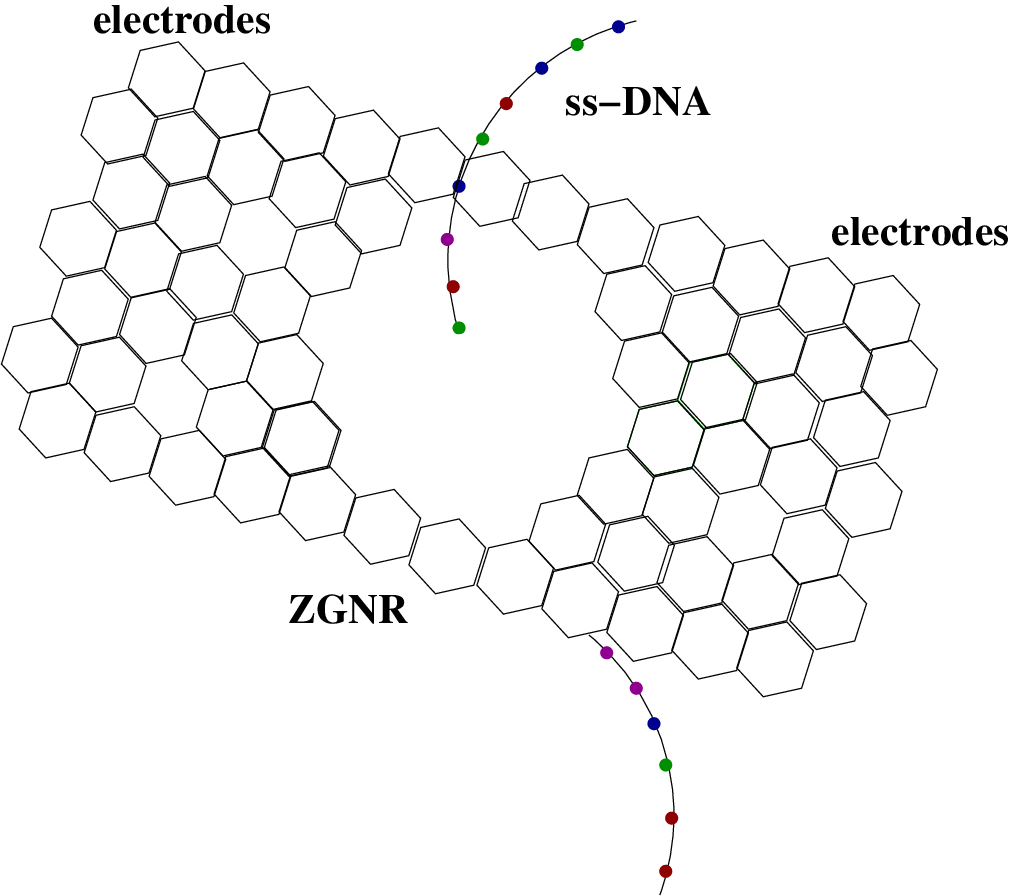}

\caption{(Color online). Schematic view of a ZGNR nanopore with a ss-DNA 
passing through it. The color dots in ss-DNA represent different nucleotides. ZGNR electrodes are shown in the figure on the both sides of the device.}
\label{fig2}
\end{figure}

 For numerical calculations, the site energies of the nucleotides are taken as their ionization 
potentials, extracted from the {\it ab-initio} calculations of Senthilkumar {\it et. al.}~\cite{senth} 
which gives: $\epsilon_G$= 8.178, $\epsilon_A$= 8.631, $\epsilon_C$= 9.722,and $\epsilon_T$= 9.464, all units are in eV. Then we shift the reference point of energy to the average of the ionization potentials of the nucleotides which is $\epsilon $=8.995 eV, and with respect to this new origin of energy the on-site energies for bases G, A, C, and T become -0.82 eV, -0.37 eV, 0.72 eV, and 0.47 eV respectively. The transport behaviour of the system would not be effected due to this shift of origin of energy. This is equivalent to some earlier reports~\cite{paez1,paez2,guo,sourav2} where the average of ionization potential was set as the backbone site-energy. In order to show the efficiency and reliability of the proposed device, we first calculate the LDOS. Fig.~\ref{fig2} shows the variation of the LDOS at nanopore site where we trap the nucleotides. We study this LDOS response of the bases as individual {\it i.e.}, we trap the four nucleotides A, T, G, C one by one inside the nanopore and study their LDOS profiles. The four nucleotides have distinct LDOS patterns with peaks around their characteristic site energies. The position and height of the peaks are different signifying their different electronic structure. The peak value for G is much smaller than others, which means the charge density at pore due to presence of G is much smaller than others. This relative difference of LDOS corresponding to the four nucleotides gives a chance to detect them using ARPES technique by trapping them one at a time inside the nanopore. As the behaviour of LDOS is mainly controlled by the nucleotides and the effects of backbones are not very important~\cite{he}, the LDOS measurement technique could be helpful for identification purposes of the nucleotides.

%%%%%%%%%%%%%%%%%%%%%%%%%%%%%%%% fig3
\begin{figure*}[ht]
  \centering
  \begin{tabular}{cc}

    \includegraphics[width=63mm,height=48mm]{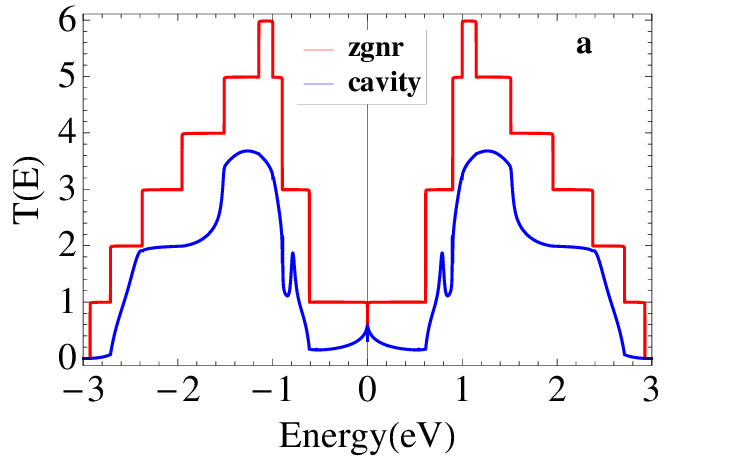}&
   
    \includegraphics[width=63mm,height=48mm]{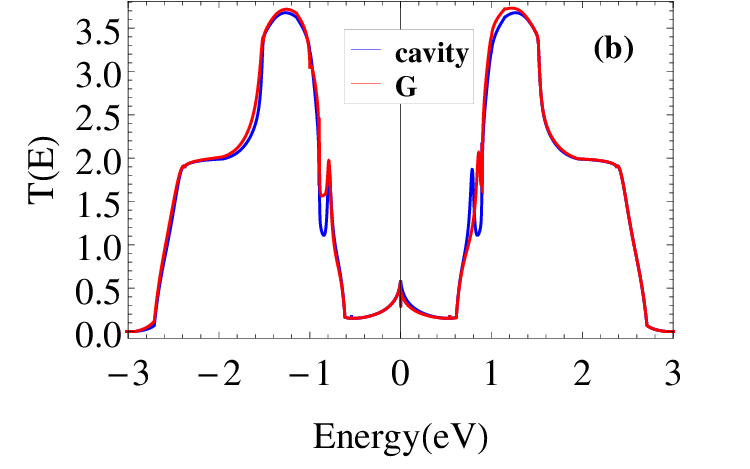}\\

    \includegraphics[width=63mm,height=48mm]{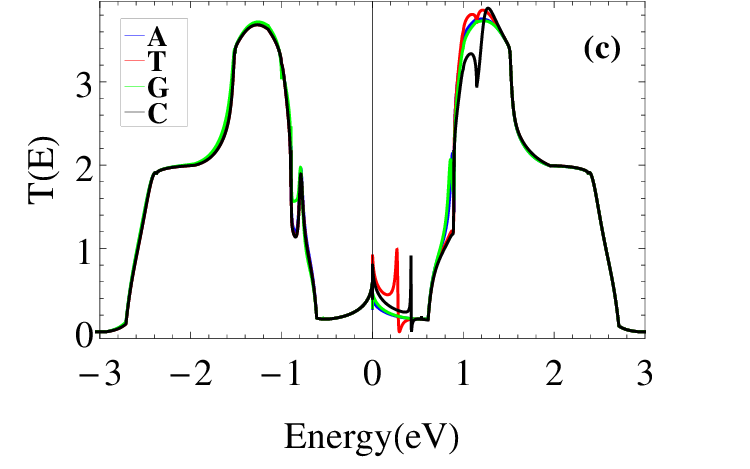}&

    \includegraphics[width=63mm,height=48mm]{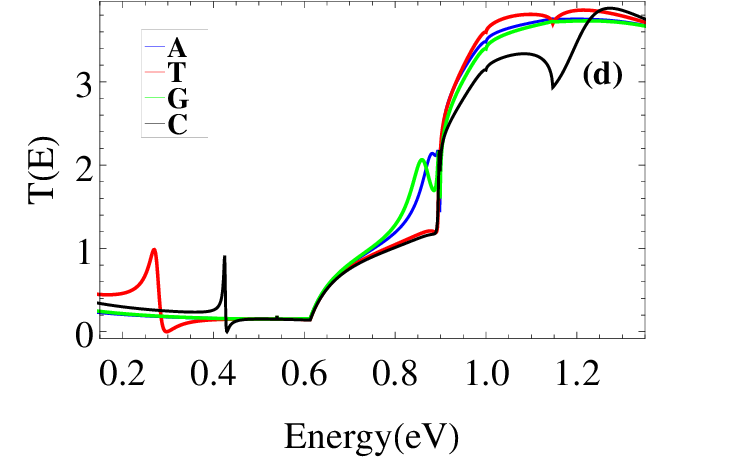}\\
    
 \end{tabular}
\caption{(Color online). Transmission probability T(E) with energy for different cases. (a) Comparison between intrinsic zgnr and zgnr-nanopore, two-sublattice symmetry being preserved. (b) Curves for a bare nanopore and G-nanopore: breaking of two-sublattice symmetry of graphene. (c) Characteristic features of each of the four different nucleotides trapped 
inside the nanopore. (d) Enlarged view of a section of the plot (c) for better visualization.}
\label{fig3}
\end{figure*}

 Fig.~\ref{fig3} shows the transmission spectra of the zgnr-nanopore device for different cases. The 
hopping parameter between two identical nucleotides in the DNA chain is taken as $t_{i,i+1}$=0.35 eV and for different bases $t_{i,i+1}$=0.17 eV~\cite{guo,sourav2}. The coupling parameter between the boundary sites of nanopore and the nucleotides is taken as $\tau'$=0.2 eV. For both the intrinsic zgnr and zgnr-nanopore the transmission profiles are symmetric with respect to zero of energy (Fig.~[3a]) and it gets distorted as we insert a nucleotide (Guanine) inside the pore (Fig.~[3b]). This is due to the fact that while creating the nanopore we have preserved the bipartite two-sublattice symmetry of graphene. As soon as a nucleotide is inserted into the nanopore this two-sublattice symmetry breaks down and the transmission spectrum becomes asymmetric with respect to E=0. Fig.~[3b] depicts the relative changes in the transmission probabilities for a bare nanopore and a G-nanopore (nucleotide Guanine trapped inside the pore). There are two distinct peaks in the transmission profile for bare nanopore placed symmetrically at +ve and -ve energies and in presence of Guanine base these peaks still remain but get distorted to some extent in asymmetric manner showing sensitivity of our device to the presence of the biomolecule. This feature of bipartite two-sublattice symmetry breaking of graphene during ss-DNA sequencing has never been reported earlier.

In Fig.~[3c] we present the transmission profiles of the nanopore device inserting each of the four nucleotides into the nanopore. The difference between the four transmission spectra are much prominent in the +ve energy region. In Fig.~[3d] we zoom in the curves of Fig.~[3c] in the energy interval 0 to 1.4 eV for better visualization. Thymine and Cytosine bases are clearly differentiable in the range 0.2 to 0.5 eV with two distinct characteristic peaks in their transmission profiles. In the midway between 0.8 to 1.0 eV Adenine (A) and Guanine (G) become differentiable. After that, at higher energy values nucleotides become less differentiable as we go far away from their characteristic site-energies. 

%%%%%%%%%%%%%%%%%%%%%%%%%%%%%%%%%%% fig4
\begin{figure}[ht]
  \centering
  \begin{tabular}{c}

    \includegraphics[width=60mm,height=45mm]{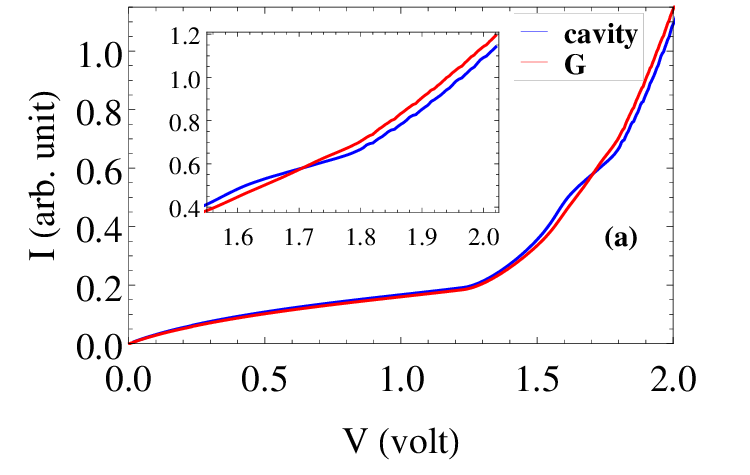}\\
   
    \includegraphics[width=60mm,height=45mm]{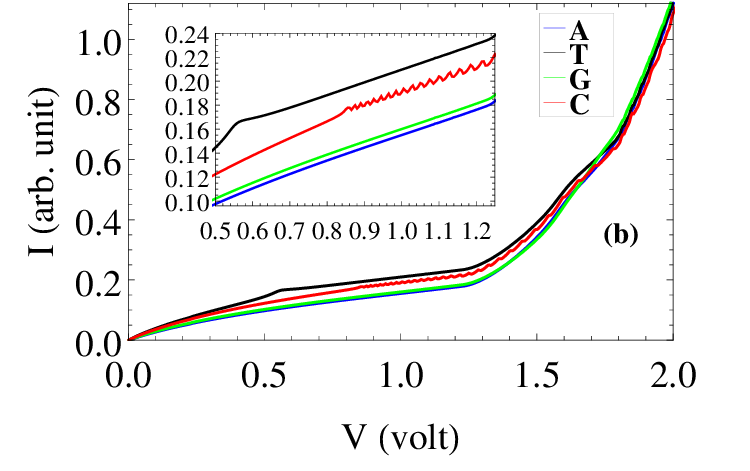}\\

 \end{tabular}
\caption{(Color online). Current - Voltage response of the active nanopore device for different cases. (a) Curves for a bare nanopore and G-nanopore. (b) Characteristic current outputs for each of the four different nucleotides trapped inside the nanopore. The insets show the enlarged view at important voltage ranges.}
\label{fig4}
\end{figure} 

Fig.~[4a] shows the I-V responses of the device for bare-nanopore and G-nanopore. The effect of the nucleobase becomes prominent at considerable bias (inset of Fig.~[4a]). Fig.[4b] shows the I-V responses of the proposed nanopore device in presence of the four nucleotides. All the four bases are distinguishable from the I-V response for most of the applied voltage range, though curves are found to be maximally separated in the range from 0.5 to 1.2 volt (inset of Fig.~[4b]). The mutual separation of the curves for T, G, C are quite large whereas that between A and G is much smaller, this might be due to the fact that both A and G belong to the same purine group. 

%%%%%%%%%%%%%%%%%%%%%%%%%%%%%%%%%%% fig5
\begin{figure}[ht]
  \centering
  
    \includegraphics[width=61mm,height=45mm]{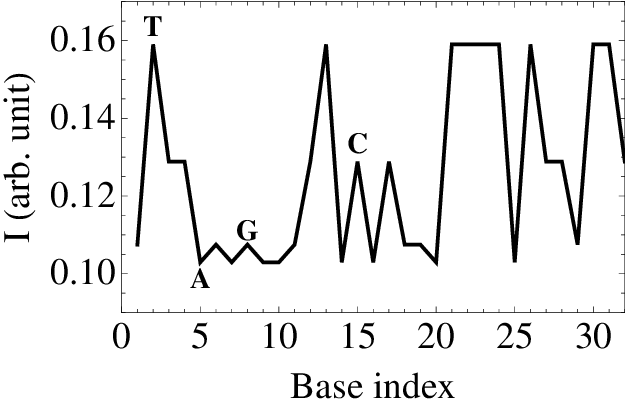}
    
\caption{Stop and go translocation of a Random ATGC ss-DNA chain through the nanopore, while bias across the device is fixed to a specific value (0.54 Volt) which gives maximum separation in current responses for different nucleotides. We record the characteristic current output for the nucleotides as they translocate through the nanopore. Respective nucleotides are represented within the figure by their usual symbols A, T, G and C adjacent to their characteristic current outputs. The sequence of the DNA chain is GTCCAGAGAAGCTACACGGATTTTATCCGTTC.} 

\label{fig5}
\end{figure}

\begin{figure}[ht]
  \centering
  
    \includegraphics[width=61mm,height=45mm]{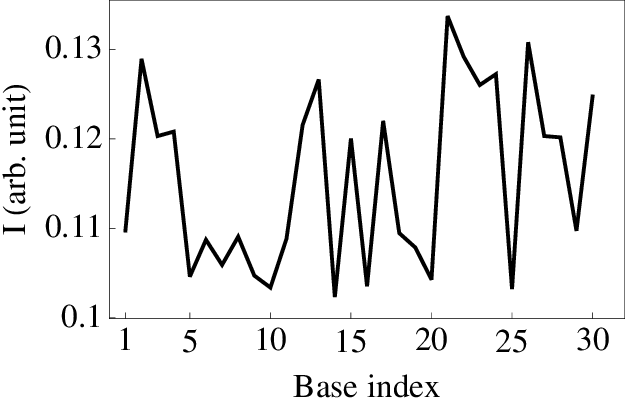}
    
\caption{Stop and go translocation of a 30-base long random ATGC ss-DNA 
chain through the nanopore, incorporating the fluctuations of the nucleotides within the nanopore during translocation while bias across the device is fixed at 0.54 Volt. We record the characteristic current output for the nucleotides as they translocate through the nanopore. The sequence of the DNA chain is GTCCAGAGAAGCTACACGGATTTTATCCGT.} 

\label{fig6}
\end{figure}

 Finally we present the actual sequencing action of the nanopore device. In Fig.~[5] we present the 
current signal of the device as we translocate a random ss-DNA through nanopore. We take a Random ATGC chain, translocate it through the nanopore and record the characteristic current signals of each nucleotides. During this translocation the bias is kept fixed at 0.54 Volt, the voltage that gives maximum separation between the characteristic currents of the four nucleotides. This can be found out by studying the current responses of the device while each and every nucleotide is trapped into the nanopore at static condition (Fig.~\ref{fig4}). The relative separation of the signals between pyrimidine and purine groups are much larger than that between two bases of the same group. The separation between Adenine (A) and Thymine (T) is maximum, while that between Adenine (A) and Guanine (G) is much smaller. The maximum and minimum values of percentage separation ($(I_{max}-I_{min})/I_{min}$) of the signals are 54.45$\%$ and 4.42$\%$ respectively, which implies that the signals corresponding to the four nucleotides could be differentiated without much ambiguity.
%more reliability than before~\cite{saha,postma,lagerq,pathak,prasong1,chien}. 
It is worthwhile to mention that though we have presented current in arbitrary unit, but if we put the numerical values of the various constants {\it e.g.}, h, $\hbar$ and e, it turns out that the currents are of the order of 10$\mu$A, being much higher than earlier reports as well as much greater than the noise level of this type of device which being of the order of nA~\cite{saha}. These high current output promises that the device can perform sequential detection even under environmental fluctuations without much uncertainty (see Fig.~\ref{fig6}). A recent report on dynamical effects of environment on the conductance of graphene nanopore devices~\cite{ralph1}, shows that fluctuations of the nucleotides inside the nanopore may change the conductance of the devices relying on tunneling mechanism, though they conclude that these effects would not be very important for the devices which rely on transverse conductance with larger transmission probabilities. Whereas a study by Krems {\it et al}~\cite{krems} in 2009 dealing with different types of noises which may occur in actual sequencing experiments showed that these environmental effects do not strongly influence the current distributions and working efficiency of these devices. Though based upon these reports we can say that the overall sensitivity of the device won't be hampered too much even in presence of environmental effects, as the proposed device relies on transverse conductance and produces greater current output, but there will always be sources of noises in actual experimental condition due to environmental fluctuations, presence of water~\cite{ralph2} and counterions which can affect device operation. Being said that, it is also important to note that the results given in this work is open to improvements by functionalization of the edge atoms of the nanopore which can significantly enhance nucleobase-pore interaction, thus reducing the structural noise by enhancing the graphene-nucleobase electronic coupling~\cite{he1,garaj}. Different types of groups can be used for functionalization (e.g., hydroxide~\cite{jeong}, amine or nitrogen~\cite{saha}) to provide custom made solution to overcome noise in electrical DNA sequencing techniques. 

%%%%%%%%%%%%%%%%%%%%%%%%%%%%%%%%%%% fig6

In Fig.~\ref{fig6} we present the variation of current responses of the device for a translocating ss-DNA chain as we incorporate the fluctuations of the nucleotides inside the nanopore during translocation. Due to these fluctuations DNA nucleotides will undergo several random orientations inside nanopore and it will result in changes of nucleotide-to-nanopore coupling~\cite{prasong}. We simulate these fluctuations in our model by considering the coupling between edge atoms of nanopore and the nucleotides ($\tau'$) to be randomly distributed within the range 0.01 to 0.2 eV. It is clear form the figure that due to these fluctuations the magnitude of the responses get decreased from their characteristic values (Fig.5) and there is also variation in the characteristic current amplitudes. As an example current output for G varies from 0.1078 to 0.1097 (arb. unit). For T current amplitude varies from 0.1260 to 0.1337 in arb. unit which has decreased from its characteristic value of 0.1590 (arb. unit). Similarly for C current output varies within the range 0.1200 to 0.1219, decreased from 0.1288 and for A current varies from 0.1023 to 0.1059, changed from characteristic value of 0.1029, all units are in arb. unit. From these specific values we can see that variations are not very large and they are not at all overlapping with each other, which shows that even in presence of this kind of environmental effects our device can work without much ambiguity. Even the drops in their characteristic current outputs are also small, maximum drop has been observed for T and minimum for A (in some cases current increases). The lowest value of current output achieved under these fluctuations is 0.1023 in arb. unit (that is actually 0.1023$\times$10=1.023 $\mu$A) which also shows that the device can produce higher current output even in presence of this type of fluctuations.

\section{Conclusion:} 
 
We present a tight-binding model study for single-molecule DNA sequencing using graphene-based nanopore devices. Most of the previous proposals on graphene nanopore based detection techniques rely on tunnelling currents that are mediated via energy levels introduced by the nucleotides~\cite{postma,lagerq,zikic,zhang,meunier}. On the other hand the current device relies on comparatively large transverse conduction current which gets modulated in presence of the DNA bases.  Presence of a nucleotide inside the nanopore modifies the LDOS at and around their characteristic site-energy and affects the transmission profiles only in the neighbourhood of this energy. The resulting transverse current is also distinguishable for each DNA bases within certain range of applied bias, and hence it renders graphene-nanopore system as a very efficient sequencing device for biomolecular detection. By modifying thses kind of devices with hybrid strucure~\cite{ralph3} one can further increase the resolving power. We also propose that the device needs to be characterised first in the static condition to find out at what condition it is maximally sensitive and then go for sequencing application by fixing the bias voltage across the device and translocating a ss-DNA through the nanopore, while recording the distinct current responses corresponding to the different nucleotides. With the two-sublattice symmetry breaking of graphene due to insertion DNA nucleotides, larger current output and appreciable amount of accuracy, we have been able to put forward a robust understanding for sequential detection of DNA bases that also offers detectable base-specific signatures even under environmnetal fluctuations (see Fig.~\ref{fig6}). We have also checked our results for other values of coupling parameter ($\tau'$), the base-specific signatures always remains with some change in the characteristic current amplitudes. In summary, our model calculations provide a comprehensive detail of the effects of four nucleotides in a nanopore sequencing set up that has strong potential applications.

\end{document}